\documentclass[aps,twocolumn,prl,showpacs,preprintnumbers,amsmath,amssymb,floatfix]{revtex4}
\usepackage{graphicx}
\usepackage{dcolumn}
\usepackage{bm}

\begin{document}

\preprint{APS/123-QED}

\title{Field-induced suppression of the heavy-fermion state in YbRh$_2$Si$_2$}
\author{Y. Tokiwa}
\author{P. Gegenwart}
\author{T. Radu}
\author{J. Ferstl}
\author{G. Sparn}
\author{C. Geibel}
\author{F. Steglich}
\affiliation{Max-Planck-Institute for Chemical Physics of Solids,
D-01187 Dresden, Germany}

\date{\today}

\begin{abstract}
We report DC magnetization measurements on YbRh$_2$Si$_2$ at
temperatures down to 0.04~K, magnetic fields $B\leq 11.5$~T and
under hydrostatic pressure $P\leq 1.3$~GPa. At ambient pressure
a kink at $B^{\star}$=9.9\,T indicates a new type of
field-induced transition from an itinerant to a localized
$4f$-state. This transition is different from the metamagnetic
transition observed in other heavy fermion compounds, as here
ferromagnetic rather than antiferromagnetic correlations
dominate below $B^\star$.
Hydrostatic pressure experiments reveal a clear correspondence of
$B^{\star}$ to the characteristic spin fluctuation temperature
determined from specific heat.
\end{abstract}

\pacs{71.10.HF,71.27.+a}
\maketitle The $f$-electrons in certain Lanthanide and Actinide
compounds can exhibit a dual character, i.e. localized as well
as itinerant, and the competition between both leads to the
exciting heavy fermion (HF) state. Here, the $f$-electrons
behave as local moments at temperatures above the Kondo
temperature $T_{\rm K}$, following a Curie-Weiss law in the
magnetic susceptibility, while the weak hybridization between
the $f$- and conduction electrons leads at lower temperatures
to the formation of heavy quasiparticles. An extremely large
mass enhancement up to $\sim$1000 in certain Ce-, Yb- and
U-based compounds has been observed. The application of
magnetic field destroys the HF state and can produce a sharp
metamagnetic-like transition from the itinerant to the
localized $f$-electron state which is in contrast to the smooth
cross-over in temperature variation at zero field. Such
metamagnetic-like behavior has been observed in Ce- and U-based
HF compounds such as CeRu$_2$Si$_2$, CeCu$_6$ and
UPt$_3$~\cite{Hae87,Loe93,Fri85}. The metamagnetic transition in
CeRu$_2$Si$_2$ has been studied most extensively because of the
dramatic step observed in magnetization and the relatively small
critical field of $B_M$=7.7\,T. Fermi surface properties studied
by de Haas-van Alphen effect(dHvA) are well explained by the
picture of itinerant and localized 4$f$-electrons below and
above $B_M$, respectively~\cite{Aok93,Yam93}. Reflecting the
localization of 4$f$-electrons, the Sommerfeld coefficient
$\gamma$ in CeRu$_2$Si$_2$ is strongly suppressed above the
critical field $B_M$~\cite{Meu91}. It is also noteworthy that
the transition is accompanied with a sharp step in
magnetostriction $\Delta L(B)/L$~\cite{Mig88}. The step
produces a sudden change in hybridization between 4$f$- and
conduction electrons. In the case of Yb-based compounds
YbCu$_{5-x}$Ag$_x$ shows a metamagnetic-like smooth cross-over
from valence-fluctuating state to a stable Yb$^{3+}$ state with
localized magnetic moments \cite{Tsu01}.

In this Letter we report a new type of field-induced suppression
of the HF state in the Yb-based compound YbRh$_2$Si$_2$. This
system with a Kondo temperature of about 25~K is located very
close to a quantum critical point (QCP), related to a very weak
antiferromagnetic (AF) order below $T_{\rm
N}$=70\,mK~\cite{Tro00}. A tiny critical magnetic field of
$B_c=0.06$~T, applied in the easy magnetic plane perpendicular
to the tetragonal $c$-axis, is sufficient to suppress the AF
order~\cite{Geg02}. For $B>B_c$, Landau Fermi liquid (LFL)
behavior is deduced from the electrical resistivity, described
by $\rho(T)=\rho_0+AT^2$, with the coefficient $A(B)$ diverging
towards $B_c$ indicating a field-induced QCP~\cite{Geg02,Cus03}.
Correspondingly, for $B>B_c$ the low-temperature specific heat
divided by temperature saturates and the Sommerfeld coefficient
$\gamma(B)$ decreases rapidly with increasing field, indicating
a strongly field-dependent quasiparticle mass in the LFL state.
It has been discovered by $^{29}$Si-nuclear magnetic resonance
experiments that the critical fluctuations in the field-induced
LFL state have a strong ferromagnetic (FM) component, dominating
for fields above 0.25\,T~\cite{Ish02}. These FM fluctuations
lead to a strongly enhanced magnetic susceptibility as
evidenced by a Sommerfeld-Wilson ratio of
$R\simeq14$~\cite{Geg02}.

The Gr\"{u}neisen parameter which describes the volume dependence
of the Kondo temperature, $\Gamma=-\partial\ln T_{\rm
K}/\partial\ln V$, is generically large in HF systems, reaching
values up to a few hundreds~\cite{deV90}. Assuming a correlation
between the characteristic field $B^{\star}$, necessary to
suppress the Kondo state, and $T_{\rm K}$, one would expect a
strong volume dependence of $B^{\star}$ as well. Contrary to the
Ce-case, in Yb-based systems the exchange interaction between
localized $4f$-electrons and the conduction electrons decreases
upon applying pressure, leading to a decrease of $T_{\rm K}$.
Correspondingly, a decrease of the field $B^{\star}$ with
pressure is expected. In order to search for an
itinerant-localized $4f$-electron transition in YbRh$_2$Si$_2$
under magnetic field, we have performed dc-magnetization
measurements at temperatures down to 40~mK and fields up to
11.5~T both at ambient pressure, as well as - for the first time
in any system at mK temperatures - under hydrostatic pressure up
to 1.3~GPa.

High-quality single crystals ($\rho_0=1~\mu\Omega$cm) were grown
from In-flux as described earlier~\cite{Tro00}. The DC magnetization
was measured utilizing a high-resolution capacitive Faraday
magnetometer~\cite{Sak94}. In order to determine the magnetization
under hydrostatic pressure, a miniaturized CuBe piston-cylinder
pressure cell of 6\,mm outer diameter and 3.2\,g total weight has
been designed. The piston is made from NiCrAl, a hard material with
a relatively small magnetization. The magnetization of the pressure
cell including the 6.0~mg YbRh$_2$Si$_2$ single crystal mounted on
the magnetometer, can be detected with a resolution as high as
$10^{-5}$~emu. The contribution of the sample to the total
magnetization of the sample and pressure cell is larger than 63\% in
the entire field and temperature range. The pressure is determined
by the difference between the superconducting transitions of two
small Sn samples; one placed inside the pressure-transmitting medium
(daphne oil) together with the YbRh$_2$Si$_2$ sample, the other one
outside the pressure cell. The $T_c$ values are determined using a
commercial SQUID magnetometer. In order to investigate the field
dependence of the quasiparticle mass, specific heat measurements
have been performed with the aid of a quasi-adiabatic heat-pulse
technique.

\begin{figure}[t]
\includegraphics[height=8cm,keepaspectratio]{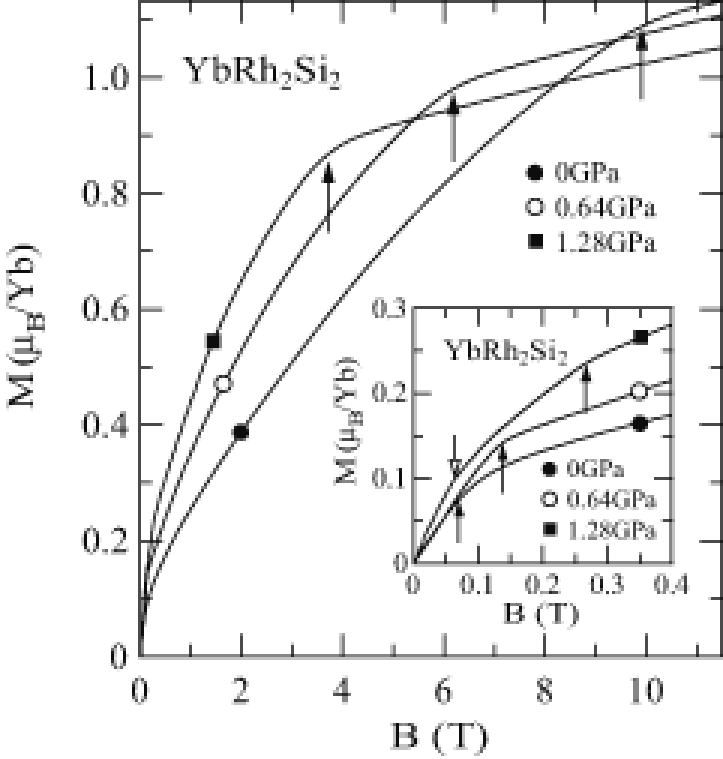}
\caption{\label{Fig1}Field dependence of the magnetization($B\perp
c$ and in units of effective moment per Yb) at differing pressures
of 0, 0.64 and 1.28\,GPa measured at 40, 40 and 60\,mK,
respectively. Arrows indicate respective values of critical field
$B^\star$ above which the 4$f$-states are localized. Inset enlarges
low-field regime. Filled and open arrows indicate critical fields
for the AF order and second transition, induced at high pressure,
see text.}
\end{figure}

Figure 1 shows the low-temperature magnetization $M(B)$ at several
different pressures. At ambient pressure the magnetization curve
reveals two kinks. The first one at very low fields (see inset)
results from the suppression of AF order at $B_c=0.06$~T. Note that
the polarization at $B_c$ amounts to $\sim 0.1~\mu_B$, only. The
remaining part of the moment is fluctuating and contributes to the
strongly enhanced Pauli-paramagnetic susceptibility~\cite{Geg02}.
The second kink at $B^{\star}$=9.9\,T, we ascribe, as discussed, to
the itinerant-localized transition of the $4f$-electrons. For
magnetic fields $B>B^{\star}$, the magnetization tends to saturate
at a value of the order of 1.2~$\mu_B/Yb$ expected for a polarized
doublet ground state. This upper kink broadens rapidly with
increasing temperature and disappears above 2\,K without shifting
its position in field. Upon applying hydrostatic pressure, the kink
shifts to $B^{\star}$=6.2 and 3.7~T at 0.64 and 1.28\,GPa,
respectively. Thus, this anomaly is very sensitive to pressure.

The pressure dependence of the AF phase transition has been
studied by electrical resistivity measurements\cite{Med02}.
Mederle {\it et al.} found an increase of $T_N$ with pressure
and the indication of a second phase transition, labeled $T_L$
inside the antiferromagnetically ordered state for pressures
above $\sim$1\,GPa. As shown in the inset of Fig.1, the
critical field $B_c$ for the AF order increases to 0.14 and
0.29\,T at pressures of 0.64 and 1.28\,GPa, respectively. The
kink at $B_c$ sharpens substantially under pressure. The
additional anomaly observed at 0.08\,T for a pressure of
$p=1.28$\,GPa indicated by the empty bracket corresponds to the
suppression of the state below $T_L$ observed by Mederle {\it
et al.} inside the antiferromagntically ordered state.

\begin{figure}[t]
\includegraphics[height=7.5cm,keepaspectratio]{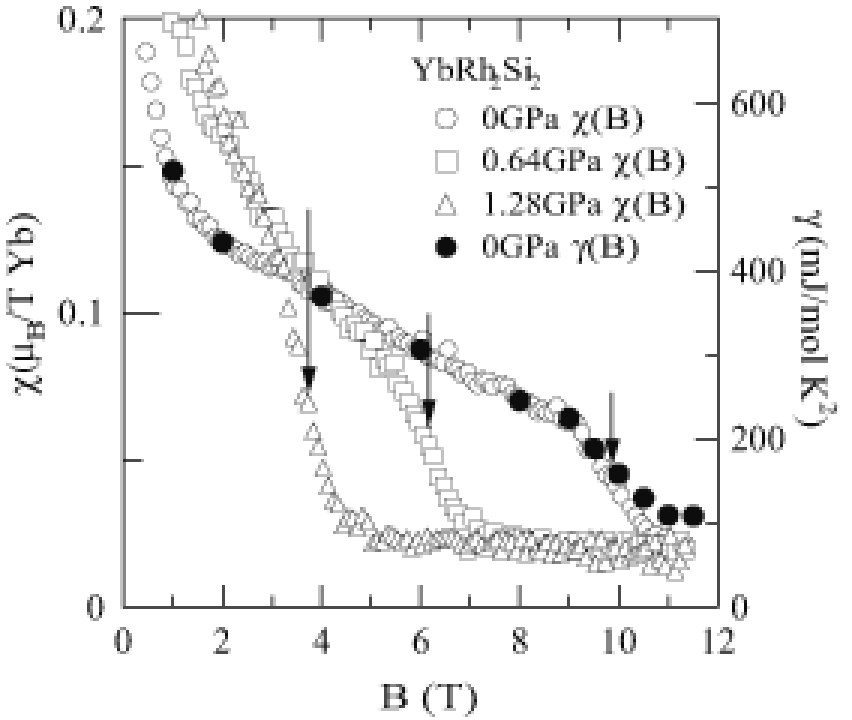}
\caption{\label{fig2}Field dependence of the differential
susceptibility $\chi=dM/dB$ at 0, 0.64 and 1.28\,GPa (left axis)
as well as Sommerfeld coefficient $\gamma$ at ambient
pressure(right axis). Arrows indicate critical field $B^*$ above
which the 4$f$-states are localized.}
\end{figure}

As shown in Figure 2, both the differential susceptibility
$\chi(B)=dM(B)/dB$ and the Sommerfeld coefficient $\gamma(B)$
show a broadened step at ${B^\star}$ that sharpens under
hydrostatic pressure. The fact that $\chi(B)\propto\gamma(B)$
for $B\leq B^{\star}$ proves a low-field LFL state of itinerant
$4f$-electrons with strongly field dependent quasiparticle
mass. Using an effective moment of $1.4 \mu_B/Yb$ \cite{Geg02},
the resulting Sommerfeld-Wilson ratio $R_W$ equals $18\pm 1$
below $B^\star$. The step-like decrease at ${B^\star}$,
indicates a large and sudden reduction of the quasiparticle
mass. A similar feature has previously been observed for the
$A$-coefficient determined from the $T^2$ contribution to the
electrical resistivity~\cite{Tok04}. Additionally, the slope of
the linear magnetostriction changes at this field from negative
to positive, suggesting the formation of completely localized
$4f$-moments in the high-field state~\cite{Tok04}. Further on,
at $P$=0 and above ${B^\star}$, $\gamma(B)$ slightly deviates
from the relation $\gamma\propto\chi$. This implies that $R_W$
decreases significantly at ${B^\star}$. Note, that $\gamma$
above ${B^\star}$ has a residual value
$\sim$100mJ/mol$\cdot$K$^2$ which is still large for a local
moment system. This might indicate residual Kondo-type
interactions persisting even above ${B^\star}$. In
CeRu$_2$Si$_2$, too, relatively large $\gamma$ values beyond
the metamagnetic transition have been found at magnetic fields
far above $B_M$($\sim$80\,mJ/mol$\cdot$K$^2$ at
20\,T)~\cite{Meu91}, although the Fermi surface properties are
in good agreement with the picture of localized 4$f$-electrons.
However, there is a distinct difference between our results on
YbRh$_2$Si$_2$ and those observed at metamagnetic transitions
in, e.g., CeRu$_2$Si$_2$ and UPt$_3$: For the latter systems the
Sommerfeld coefficient $\gamma (B)$ shows a peak at
$B_M$~\cite{Meu91,Aok98,Sug99}. The absence of a peak in $\gamma
(B)$ for YbRh$_2$Si$_2$ is related to the absence of a peak in
the susceptibility, and the origin of this difference is
discussed below.

\begin{figure}[t]
\includegraphics[height=7cm,keepaspectratio]{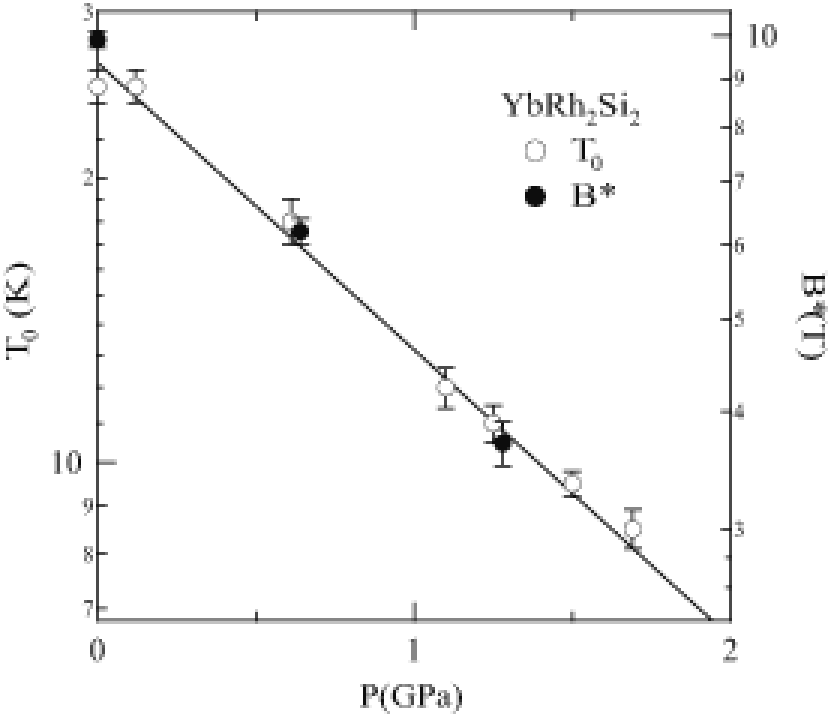}
\caption{\label{fig3}Pressure dependence of the characteristic
spin fluctuation temperature $T_0$~\cite{Med02} (left axis) and
the field ${B^\star}$ (right axis) for YbRh$_2$Si$_2$. Solid line
represents $\exp(-0.7$~GPa$^{-1}\times P)$ dependence.}
\end{figure}

Next we compare the pressure dependence of ${B^\star}$ with
that of the characteristic spin fluctuation temperature $T_0$,
estimated by fitting the zero-field low-temperature specific
heat with $C(T)/T=-D\ln(T/T_0)$. Since at ambient pressure $T_0$
matches with the single-ion Kondo temperature $T_{\rm K}$
determined from the magnetic entropy~\cite{Tro00}, the pressure
dependence of $T_0$ is assumed here to represent that of the
Kondo temperature $T_{\rm K}$. In order to obtain the pressure
dependence of $T_0$, we used specific-heat data under
hydrostatic pressure reported in~\cite{Med02}. As shown in
Fig.~3, a correlation between $T_0$ and $B^\star$ is very
probable. The exponential decrease with increasing pressure is
compatible with the Kondo temperature $T_{\rm K}\propto
\exp(-1/J_{cf}D_c(\varepsilon_F))$ being determined by the
product of the $4f$-conduction electron exchange intergral,
$J_{cf}$, and conduction electron density of states at the
Fermi energy, $D_c(\varepsilon_F)$.


Using the isothermal compressibility $\kappa_T=5.3\times
10^{-12}$~Pa$^{-1}$ \cite{Ple03} we obtain the "thermal"
Gr\"uneisen parameter $\Gamma_T=1/\kappa_T\times\partial\ln
T_{\rm K}/\partial P =-132\pm 6$. The "magnetic" Gr\"uneisen
parameter, derived from the pressure dependence of the
characteristic field $B^\star$,
$\Gamma_B=1/\kappa_T\times\partial\ln B^\star/\partial P$,
equals $\Gamma_T$ because of the same slope for $T_0$ and
${B^\star}$ in their pressure dependences. This resembles the
case of CeRu$_2$Si$_2$ for which $\Gamma_B$, determined from the
pressure dependence of the metamagnetic transition equals
$\Gamma_T$ as well. One important difference of YbRh$_2$Si$_2$
compared to CeRu$_2$Si$_2$ is, however, the strongly enhanced
Sommerfeld-Wilson ratio in the former system. A systematic
comparison of thermal and magnetic Gr\"{u}neisen parameters for
various systems has been made by A.~B.~Kaiser and P.~Fulde
~\cite{Kai88}. They found that in contrast to usual metals and
HF systems, for strongly enhanced Pauli paramagnets with a
Sommerfeld-Wilson ratio $R_W\gg 1$ the magnetic Gr\"uneisen
parameter is much larger than the thermal one. The enhancement
of the magnetic compared to the thermal Gr\"uneisen parameters
in these systems results from the strong volume dependence of
$R_W$ \cite{Kai88}. In this respect YbRh$_2$Si$_2$ is different
to nearly ferromagnetic metals, as the observed scaling
behavior between $T_{\rm K}$ and ${B^\star}$ indicates
$\Gamma_B\simeq\Gamma_T$. Probably, the pressure dependence of
$R_W$ is small compared to that of $T_{\rm K}$. Indeed a
similar $R_W$ value has been observed in
YbRh$_2$(Si$_{0.95}$Ge$_{0.05}$)$_2$ \cite{Geg05}.

\begin{figure}[]
\includegraphics[height=7cm,keepaspectratio]{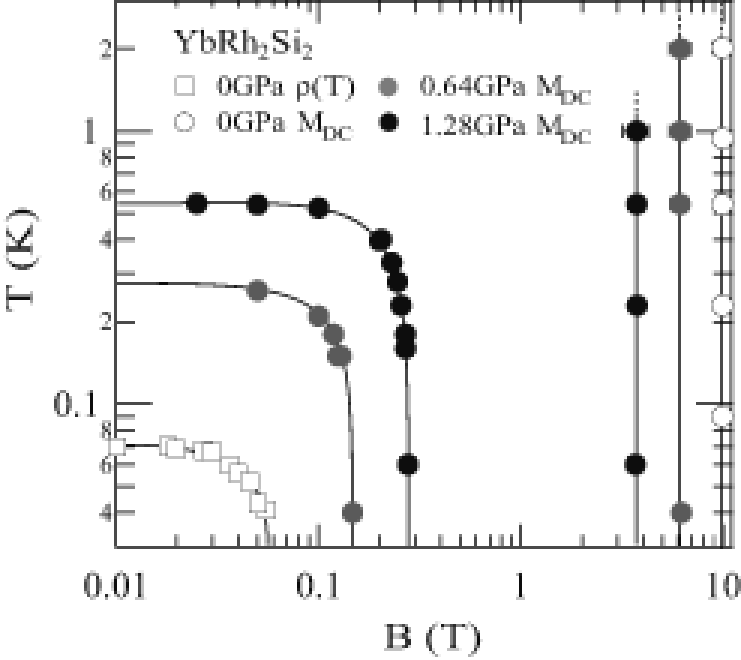}
\caption{\label{fig5}Temperature-magnetic field phase diagram of
YbRh$_2$Si$_2$ for $B\perp$ c as $\log T vs \log B$. White, gray
and black symbols indicate points at 0, 0.64 and 1.28\,GPa,
respectively. Circles and squares are data points from
magnetization and resistivity measurements, respectively.
Lines are guides to the eye.
}
\end{figure}

Our results are summarized in the $T-B$ phase diagram displayed
in Fig.4. Here the phase boundary of the AF state at low fields
has been determined from kinks in both constant temperature and
constant field scans, respectively. Zero-field extrapolations of
the phase boundaries agree well with the $T_{\rm N}(P)$ results
of electrical resistivity measurements under hydrostatic
pressure~\cite{Med02}. In all $M(B)$ measurements under
pressure, the field $B^{\star}$ at which the kink occurs is
independent of temperature. The anomaly broadens rapidly with
increasing temperature, and the kink disappears around 2~K at
ambient pressure and 1~K at the highest pressure of this study.
Thus, the transition occurs only in the coherent regime at $T\ll
T_{\rm K}(P)$.



We discuss the peculiar anomaly at ${B^\star}$ by comparing it
with the metamagnetic-like transition in HF compounds such as
CeRu$_2$Si$_2$ and CeCu$_6$. Both these compounds as well as
UPt$_3$ have been reported to exhibit AF short-range
correlations at low temperatures~\cite{Mig88_2,Reg87,Aep87}. In
CeRu$_2$Si$_2$ the AF correlations show a step-like decrease at
$B_M$ and disappear at higher fields. Therefore metamagnetism
in these HF compounds can be interpreted as a transition from
an "almost" antiferromagnetically ordered HF phase below $B_M$
to a ferromagnetically polarized localized $4f$-state without AF
correlations. It seems reasonable to suppose that the intensity
of AF correlations is related to the strength of the
metamagnetic-like behavior. The weaker peak in $\chi(B)$ at
$B_M$ for CeCu$_6$ is consistent with a smaller intensity of AF
correlations compared to that in CeRu$_2$Si$_2$. In
YbRh$_2$Si$_2$, on the other hand, AF correlations have been
found to persist only in close vicinity to the critical field
$B_c$, and the field-induced LFL state for $B>B_c$ is dominated
by strong ferromagnetic fluctuations~\cite{Ish02}. Their
polarization with increasing $B$ causes the large magnetization
already well below the transition at $B^\star$.


In CeRu$_2$Si$_2$, the strength of the AF correlations and the
Kondo interaction are comparable and thus the strong reduction of
the former and the localization of the $4f$-electrons happen at
the same field. Because of the very weak RKKY intersite
interaction in YbRh$_2$Si$_2$, evidenced by the very low ordering
temperature $T_N$, the AF correlations cannot persist at fields
needed to destroy the Kondo interaction in this system. Thus, in
contrast to the metamagnetic like behavior in other HF systems,
resulting from the cross-over from an AF correlated itinerant to a
ferromagnetically polarized localized $4f$-moment state, the
$B^\star$ anomaly in YbRh$_2$Si$_2$ may result from an itinerant
to localized transition with FM polarization at both sides.
It is worth noting that the relatively small critical field
${B^\star}$=9.9\,T is suitable for dHvA experiments to study
Fermi surface properties below and above ${B^\star}$. These
experiments will provide crucial evidence for the localization
of the 4$f$-electrons.



To conclude, a new type of field-induced suppression of the HF
state has been discovered by low-temperature magnetization
measurements on YbRh$_2$Si$_2$. At ambient pressure, we have
observed a broadened transition at $B^\star=9.9$~T which is
accompanied by a decrease of the quasiparticle mass. The use of
a miniaturized hydrostatic pressure cell for low-$T$
magnetization experiments has revealed a clear one-to-one
correlation between the transition field $B^\star$ and the
Kondo temperature. Both are strongly pressure dependent with a
Gr\"uneisen parameter of about $-130$. Strong ferromagnetic
fluctuations present in the HF state cause the unique
difference of YbRh$_2$Si$_2$ compared to all other HF systems.


We gratefully acknowledge technical support and useful advises
for the use of the hydrostatic pressure cell from C.~Klausnitzer
and T.~Nakanishi.

\end{document}